\documentclass{article}
\usepackage{emulateapj,pstricks,apjfonts}
\usepackage{lscape}
\usepackage{natbib}

\def\pks{PKS~2155-304}
\begin{document}

\submitted{Accepted version. To appear in ApJ main journal}

\lefthead{A {\em XMM-Newton} RGS view of PKS~2155-304}
\righthead{Cagnoni et al.}
\title{A view of PKS~2155-304 with {\em XMM-Newton} Reflection Grating Spectrometers}

\author{I. Cagnoni\altaffilmark{1}, F. Nicastro\altaffilmark{2}, L. Maraschi\altaffilmark{3}, A. Treves\altaffilmark{1} and F. Tavecchio\altaffilmark{3}}
\affil{$^1$ Dipartimento di Scienze, Universit\`a dell'Insubria, Via Valleggio 11, Como, I-22100, Italy}
\affil{$^2$ SAO, 60 Garden Street, 02138, Cambridge, MA, USA}
\affil{$^3$ Osservatorio Astronomico di Brera, Via Brera 28, Milano, I-20121, Italy}

\footnote{email: Ilaria Cagnoni: ilaria.cagnoni@uninsubria.it}

\begin{abstract}
We present the high resolution X-ray spectrum of the BL Lac object PKS~2155-304  
taken with the RGS units onboard {\em XMM-Newton} in  November 2000. 
We detect a  O~{\sc vii} K$\alpha$  resonant absorption 
line from warm/hot local gas at 21.59 \AA \/ ($\sim 4.5$~$\sigma$ detection). The line profile is possibly double peaked.
 We  do not confirm the strong 20.02 \AA \/ absorption  line seen with {\em Chandra} 
and interpreted as $z \sim 0.05$ O{\sc viii} K$\alpha$.
A $3$~$\sigma$ upper limit of 14 m\AA \/ on the equivalent width is set.
We also detect the $\sim 23.5$~\AA \/  interstellar O{\sc i} 
1s$\rightarrow$2p  line and derive a 
factor $\leq 1.5$ subsolar O/H ratio 
in the ISM along \pks \/ line of sight. 
\end{abstract}


\section{Introduction}
The observed baryon density at $z > 2$ \citep[e.g.][]{rau98,wei97} agrees 
well with Standard Big-Bang nucleosynthesis predictions, 
when combined with observed light-element ratios \citep{bur98}. 
At lower redshift, however, less than $1/3$ of the baryons observed 
at $z > 2$ have been detected so far \citep{fuk98}.
According to simulations for the formation of structures in the Universe,
most of such baryons would be located, at the present epoch, 
in low density intergalactic gas, which has been shock-heated to temperatures 
of $\sim 10^{5} - 10^{7}$~K 
\citep[i.e. the warm-hot intergalactic medium WHIM, e.g.][]{hel98}.
The most efficient way to detect the presence of WHIM is through resonant 
absorption lines from highly ionized metals 
(e.g. O{\sc vi}, O{\sc vii}, O{\sc viii}, Ne~{\sc ix})
imprinted in the far UV (FUV) and soft X-ray spectra of background sources 
\citep{ald94, mul96, hel98, per98, fan00}. 
However, severe instrumental limitations have hampered so far the  
detection of conspicuous amount of WHIM. 
The Far Ultraviolet Spectroscopic Explorer ({\em FUSE}) allows the OVI doublet 
($\lambda = 1031.93$ and 
$\lambda=1037.62$ \AA ) to be detected only up to z$\sim 0.2$. 
More importantly OVI dominates the relative abundance distribution of O 
in shock-heated gas in pure collisional ionization equilibrium, only 
at relatively low temperatures, $T\sim (1-5) \times 10^{5}$~K, 
and so it tracks just the low-temperature tail of the WHIM distribution 
\citep[e.g.][]{fan00}. Soft X-rays are far more 
promising. The strongest resonant transitions from O{\sc vii}, Ne{\sc ix} (both 
helium-like ions, and therefore highly stable) and O{\sc viii}, all fall in the soft 
X-ray band. These ions dominate the O and Ne relative abundance 
distributions of both collisionaly ionized gas, and mixed, photoionized 
and collisionaly ionized gas, over a broad range of temperatures, 
between $T \sim 5\times 10^5$ K and $T \sim 10^7$ K, where the WHIM 
distribution peaks \citep{dav01,fan00}.
The major fraction of the  WHIM should then be visible in the X-rays. 
However, despite the large relative abundances of ``X-ray'' ions in the 
WHIM, the current sensitivity and resolution of X-ray spectrometers 
has allowed so far only the strongest (EW$\geq 10$ m\AA) of these systems 
to be detected, and only against spectra of very bright background sources 
(Nicastro et al. 2002; Mathur, Weimberg \& Chen 2003;  
Fang et al. 2002; Fang, Weimberg \& Canizares 2003; 
with {\em Chandra} and Rasmussen, Kahn \& Paerels 2003 with {\em XMM-Newton};
for a recent review see Paerels \& Kahn 2003).

Three out of these six detections have been made against 
the spectrum of the bright ($F_{\rm 2-10 keV} \simeq 2 \times 10^{-11} - 5 \times 
10^{-10}$ erg cm$^{-2}$ s$^{-1}$), nearby  
\citep[z=0.116]{fal93}, blazar PKS~2155-304 \citep{nic02, fan02}. 
A high quality ($\geq 700$ counts per resolution 
element at the relevant wavelengths) {\em Chandra} (HRCS/LETG) 
observation of PKS~2155-304, revealed 
the existence of O{\sc vii}, O{\sc viii} and Ne{\sc ix} absorption lines 
at $z \simeq 0$,
identified with a  WHIM system, pervading our Local Group \citep{nic02}. 
Lower quality {\em Chandra} (ACIS/LETG) spectra of PKS~2155-304 confirmed 
the above findings \citep{fan02}, and also show a line at 20.02 \AA, that 
\citet{fan02} identify as the O{\sc viii} WHIM counterpart of a 
known Ly$\alpha$ and O{\sc vi} 
system at $z \sim 0.05$, where a concentration of 
galaxies is seen \citep[][and references therein]{mar88,shu98,shu03}. 
However it is not clear why such a feature is not detected 
in the higher quality HRCS/LETG spectrum published by \citet{nic02}.
Additional data are needed to clarify this issue. 

In this paper we present the analysis of the high resolution {\em 
XMM-Newton} Reflection Grating Spectrometers (RGS hereinafter) spectra of \pks .\\
The structure of the paper is as follows: 
in \S~\ref{sec_obs}  we report on the  RGS data reduction and analysis.
The spectral fitting and comparison with previous measurements is given in
  \S~\ref{sec_fits}.
The discussion of the detected features is presented in \S~\ref{sec_disc}.

\section{Observations and data reduction}
\label{sec_obs}
\pks \/ was observed by {\em XMM-Newton} in revolution 174 (Nov 19-21, 2000) 
in two short and two long pointings described in Table~\ref{tab1}.
{\em XMM-Newton} is equipped with three coaligned X-ray telescopes and 
provides images over a $30^{\prime}$ field of view with moderate
spectral resolution using the European Photon
Imaging Camera (EPIC), which consists of two MOS and one PN CCD arrays.
High-resolution spectral information is provided by the Reflection
 Grating Spectrometers (RGS-1 and RGS-2) that deflect half of the beam of two 
of the three X-ray telescopes.
Due to a failure in the read-out electronics of two RGS CCDs 
(CCD 4 in RGS-2 and CCD 7 in RGS-1) the total effective area is reduced by 
a factor 2 in the wavelength bands originally covered by these CCDs 
(10.5--14.0 \AA\ and 20.1--23.9 \AA). 
 The observatory also has a coaligned 30~cm optical/UV telescope, i.e.
 the Optical Monitor (OM). 
In this paper we  concentrate on the RGS spectra of \pks \/
 ($\Delta E / E$ from 100 to 500, FWHM, in the energy range 0.33--2.5 keV - 
5--38 \AA), the first and second order spectra of which are clearly visible in both RGS. 
EPIC and OM data are discussed in 
\citet{mar02} and Maraschi et al. (2003, in preparation).

We reprocessed the data using the  {\em Science Analysis System} (SAS)
software version 5.3.0 and the calibration files as of May 31st 2002.
Since the wavelength calibration of XMM grating spectra depends strongly 
on the position of the 0th order, we used the VLBI position as centroid 
of the 0th order source \citep{ma98}. 
Extraction regions, for source and background, were chosen to be, 
respectively, within the 95\% and outside the 98\% of the
PSF.
Since XMM-RGS observations can be affected by high particle background 
periods caused by solar activity, we extracted RGS1 and RGS2 
background lightcurves from CCD-9, the closest to the optical axis of the 
telescope and therefore the most affected by background flares. 
We excluded all the time intervals  for which the background count rate
was higher than $10^{-3}$ count s$^{-1}$ and report the net 
 exposures  in Table~\ref{tab1}.
Note that the short observation at the end of the run (Obs. Id. 0080940501) 
is completely affected by background flares and will not be further 
considered in this paper, which discusses the first three observations only.
For each observation and for each RGS unit we extracted 
first and second order source and the background spectra, for a total 
of 12 spectra, but
since second order spectra have a significantly lower number of photons, 
we will concentrate on  first order only.

\section{Spectral Fitting}
\label{sec_fits}
We modeled the continuum of each RGS-1 and  RGS-2 observation of \pks \/  
in the energy range 0.35 -- 2.0 keV using version 11.2.0 of 
XSPEC.
We adopted an absorbed power law model of the form 

\[F(E)=F(E_0) E^{-\Gamma } \exp{(-{\Sigma}_{x} N(X) {\sigma}_X) }\]

where f($E_0$) is a normalization factor at $E_0=1$~keV,
$\Gamma $ is the photon index and
the photoelectric absorption is characterized by  a column density N(X) and an
absorption cross section $\sigma_{\rm{X}}$ for each element \citep{mor83}.
The Galactic hydrogen column density  is fixed at $N_H =1.36 \times
10^{20}$~cm$^{-2}$ \citep{loc95}.
The best fit slopes and the source fluxes during the observations are 
reported in Table~\ref{tab2}.
Errors in the paper represent $1$~$\sigma$ confidence levels unless stated 
otherwise.
We note a small systematic difference between RGS-1 and RGS-2, with
RGS-1 measuring steeper slopes ($\sim +0.02$) and  ($\sim 3$\%) 
higher fluxes compared to RGS-2.
Table~\ref{tab2} shows that \pks \/ emission steepens when the source gets 
fainter, a well known behavior for this source \citep[e.g.][]{zha02}.

In order to improve the signal to noise ratio (SNR) we 
performed a simultaneous fit of the three datasets forcing the 
slope to be the same in the three observations, but leaving the 
normalization free to vary.
For display purposes we 
combined all the three first order RGS-1  spectra together
and  did the same for the RGS-2 (Fig.\ref{fig1}).

\subsection{RGS Instrumental Features}
\begin{figure}
\plottwo{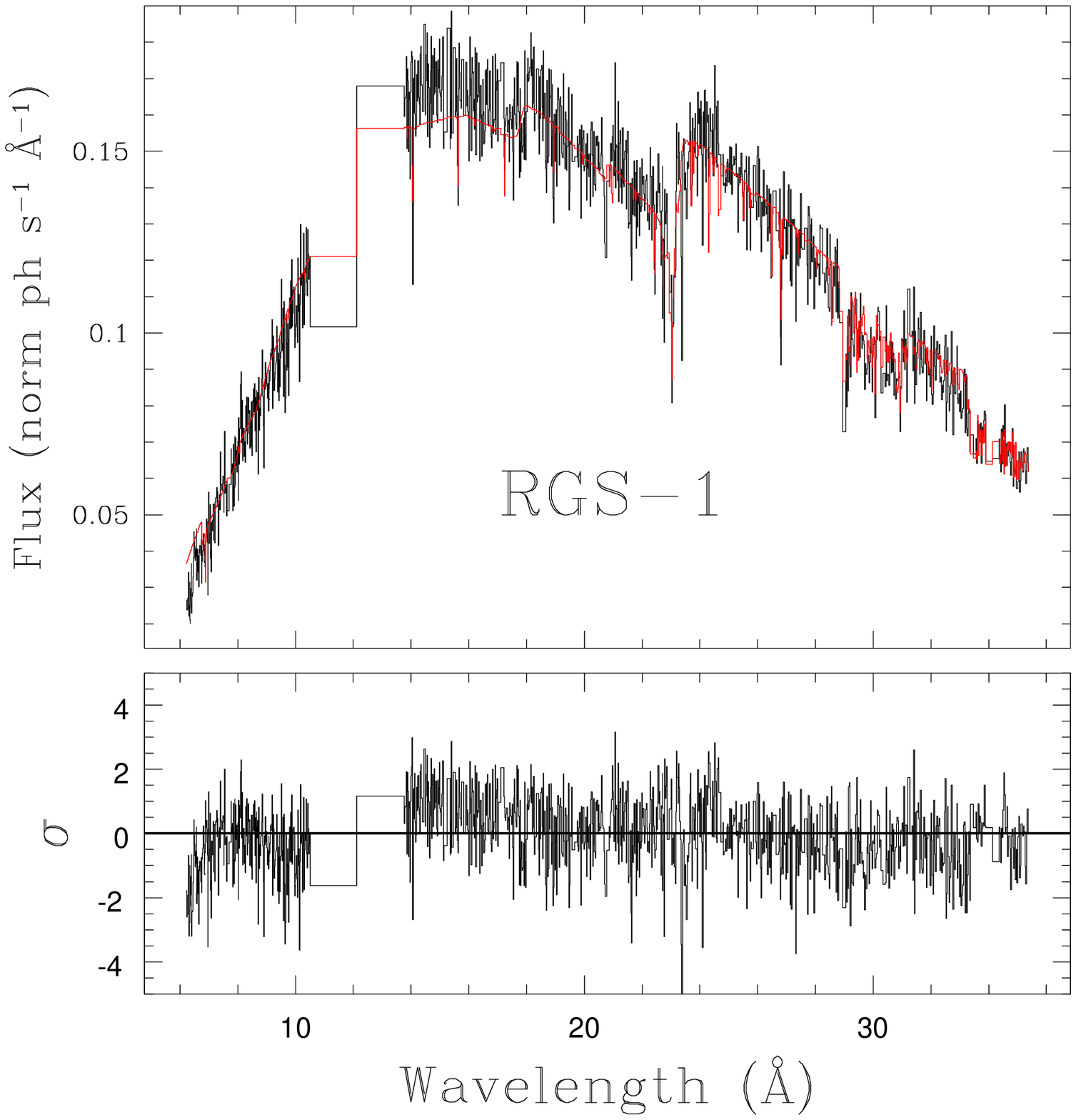}{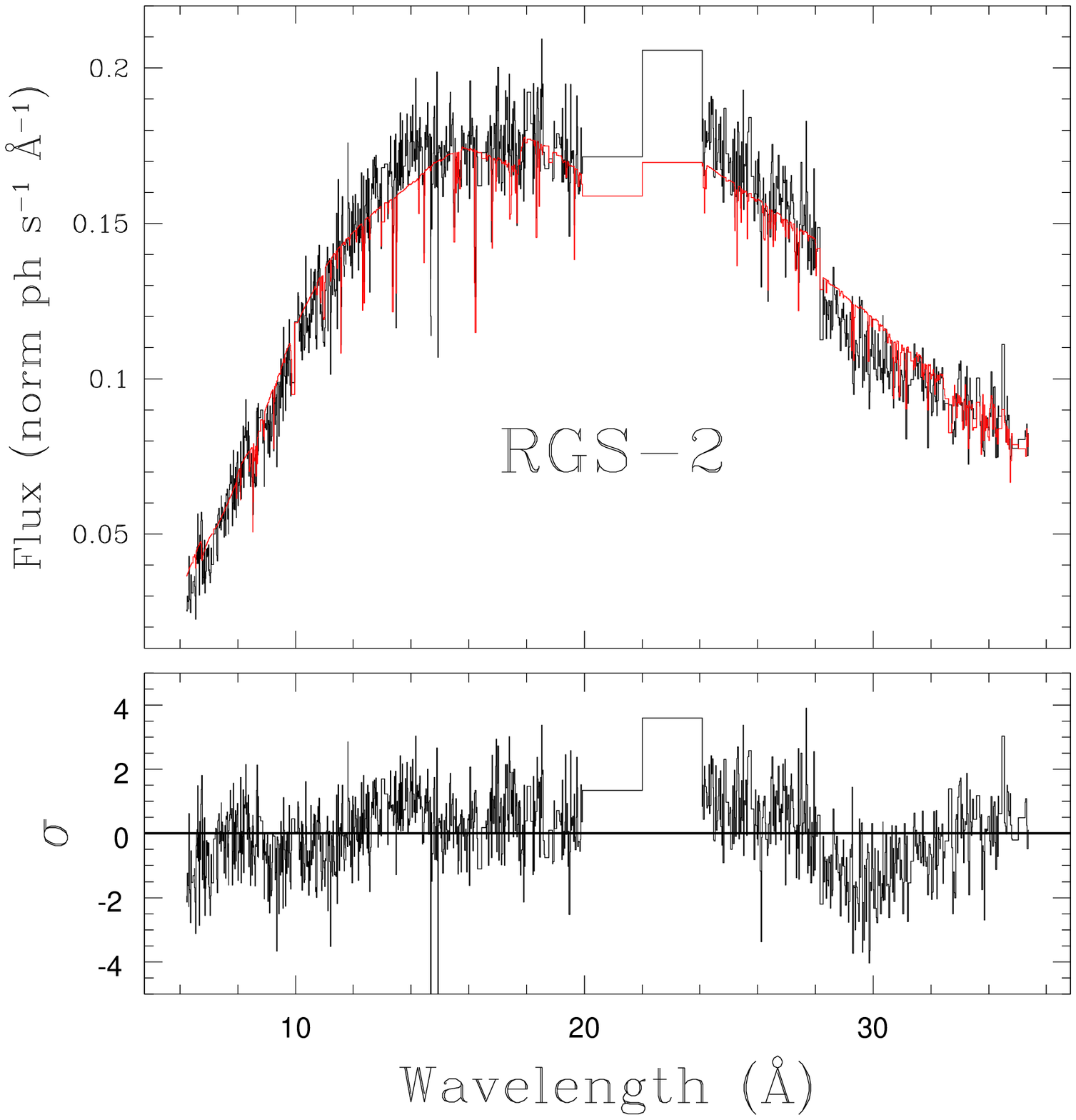}
\caption{RGS-1 (left) and RGS-2 (right) combined first order  spectra of \pks \/ and best fit absorbed power law model  folded with the instrument response. The residuals to the fits are reported in the bottom panels (see text for details). \label{fig1}}
\end{figure}

\begin{figure}
\plottwo{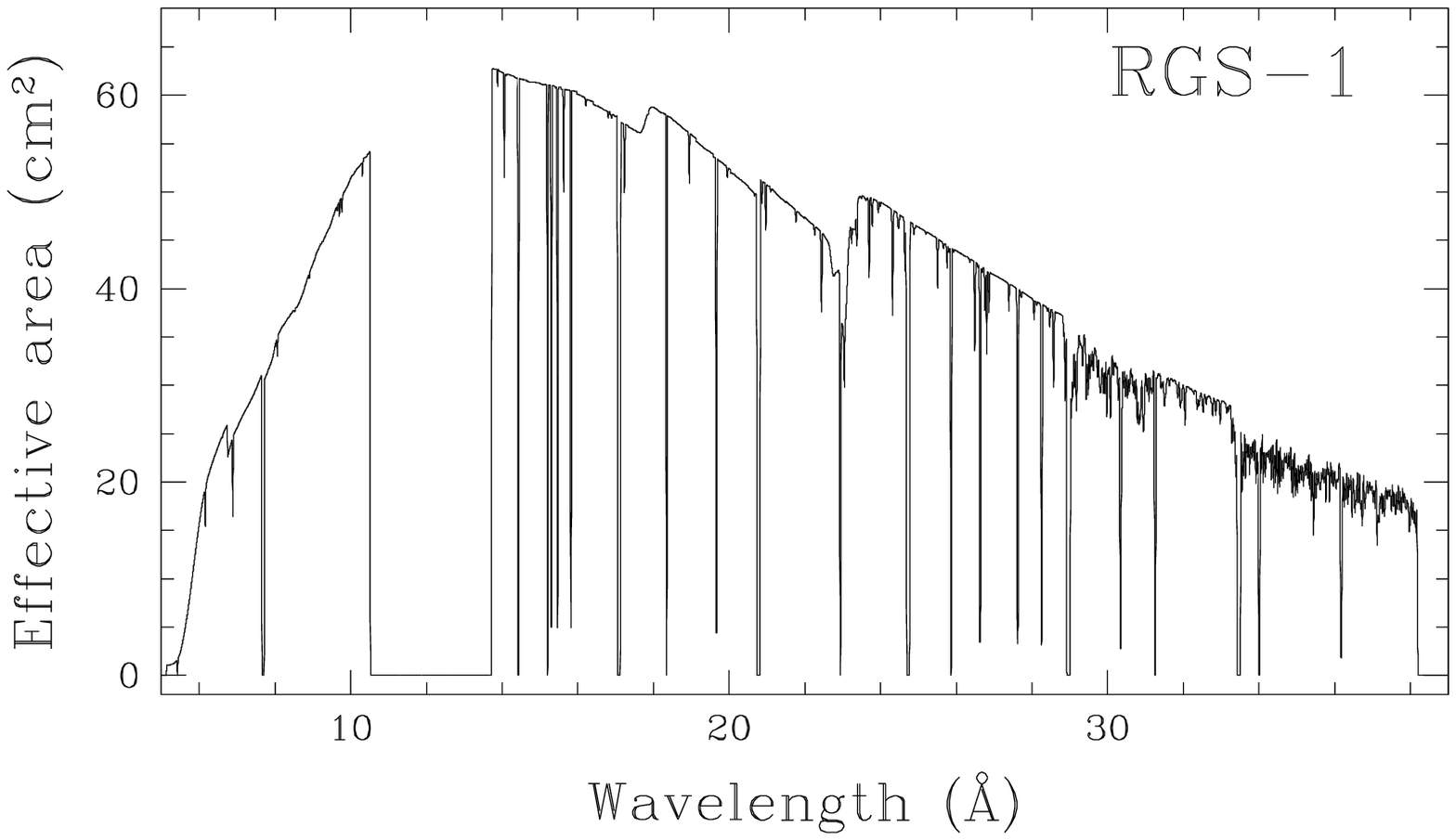}{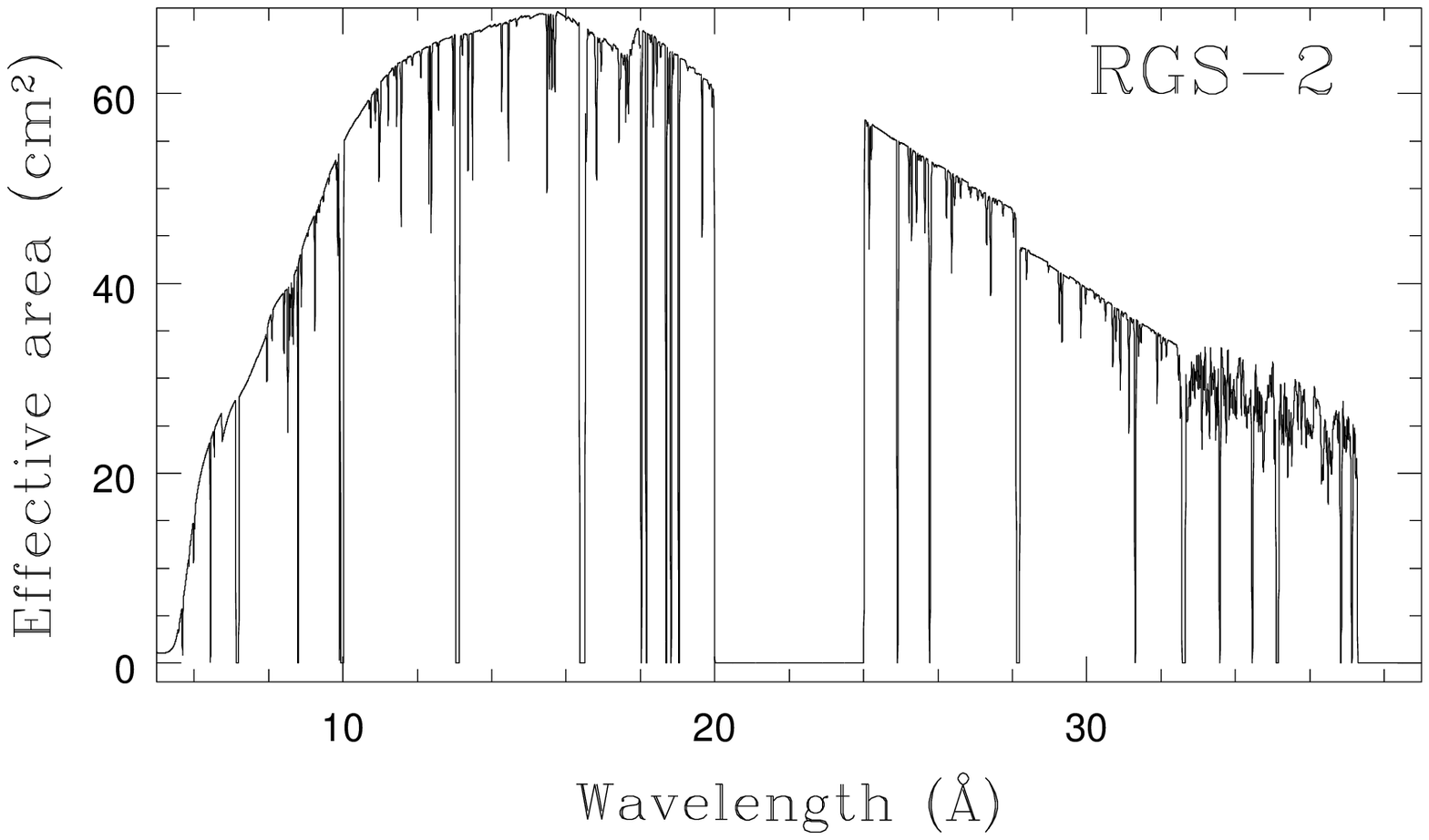}
\caption{RGS-1 (left) and RGS-2 (right) combined first order effective areas.
\label{fig2}}
\end{figure}

The RGS effective areas (see Fig.~\ref{fig2}) are complex in 
shape, and contain tens of narrow dips due to bad columns or pixels in the 
CCD detectors. The exact shape of these instrumental 
features in observed spectra depends on the grating Line Spread Function 
(LSF), which is particularly broad for reflection-gratings, due to electron 
scattering of dispersed photons \citep[i.e. $\sim 60$ m\AA \/ FWHM][, Figure~1]{pol03}.
Modeling of such narrow instrumental 
features in the {\em XMM-Newton} RGS requires, then, very accurate 
calibration measurements. Current RGS effective area 
calibration uncertainties are 
as accurate as $\sim 5-10$\% between 7 and 36 \AA \/ \citep{den02}. 
As a consequence, residual narrow instrumental
features with such relative intensities, are expected in the unfolded
  RGS spectra (both in absorption and emission).
These  are indeed observed 
in both RGS-1 and RGS-2  spectra of PKS 2155-304 (Fig.\ref{fig1}) and 
greatly hamper the detection and proper identification of real physical 
bound-bound electronic transitions from abundant metals in astrophysical 
gaseous environments. 
However, in the wavelength range we are interested in, these problems 
differently affect the RGS-1 and RGS-2.
Particularly, at redshifts between us and 
PKS~2155-304, all the strong resonant absorption lines from neutral and/or 
highly ionized O and Ne, fall in wavelength ranges (i.e. 13-14 \AA, 18-20 
\AA, 20-24 \AA) where RGS-2 spectra either do not exist (20-24 \AA, 
due to the failure of a CCD chip) or contain strong line-like shaped 
instrumental features (Fig.\ref{fig1}). 
RGS-1 spectra are, on the other hand, relatively 
instrumental-feature-free in the relevant oxygen wavelength ranges 
(Fig.\ref{fig2}). 
In the following, we present our analysis of the RGS-1 spectra 
of PKS~2155-304, and use the RGS-2 spectra only for comparison.

\subsection{The combined spectra}

We modeled the RGS-1 and RGS-2 continuum
of \pks \/ from the summed spectra
 in the energy range 0.35 -- 2.0 keV with the same absorbed 
power law model  described above
and the results are reported in Table~\ref{tab2}.

Figure~\ref{fig1} shows the best fit
 to the  \pks \/ spectra (binned over two channels) and the residuals. 
Clearly visible in the residuals of RGS-1 spectrum is 
the O-edge at $\sim 23$ \AA \/ and the O{\sc vii} K${\alpha}$ at 21.6 \AA .
The RGS-2 residuals, show instead a pronounced dip around $\sim 30$ \AA \/ and 
a deviation from a  power law below 14 \AA .
All these features, but the 21.6~\AA \/ line, are also present 
in the calibration spectrum presented in \citet{den02} RGS calibration report
 and are due to an  imperfect knowledge of the effective area.

To determine the line parameters we prefer the simultaneous fit of 
the three data sets because we can promptly see false absorptions 
caused by flickering hot pixels/columns in an observation which 
appear normal in the other observations,  and can remove the problematic
data set from the fitting procedure (e.g. O{\sc vii} K$\beta$, see Table~\ref{tab4}).

\subsection{Lines}\label{sec_lines}

To better model the underlying continuum in the proximity of 
absorption/emission features, we did not adopt the broad band 
 absorbed power law results presented above, but we divided the RGS-1 and RGS-2 
spectra into $\sim 2-3$ \AA \/ wavelength bins and 
performed a simultaneous fit with  a local absorbed power law.
In the following sections we will discuss the features with significance 
$> 2$~$\sigma$.

\subsubsection{The shift}\label{sec_shift}

To determine the absolute line position in RGS-1 we used the  23.5 \AA \/ 
interstellar O{\sc i}  1s$\rightarrow$2p 
line 
\citep[][see \S ~\ref{sec_astrolines}]{mcl98, dev03}.
The RGS-1 line position is 
$\lambda = 23.545 \pm 0.016$ \AA .
This is  shifted by 
(a) $+56$ m\AA \/ with respect to the laboratory value of 
23.489~\AA \/ measured  by \citet{kra94}, 
by (b)  $+47$ m\AA \/  with respect to the position measured in the 
{\em Chandra} LETG spectrum of the low-mass X-ray binary X0614+091
 by  \citet{pae01}
and by (c) $+36$ m\AA \/ with respect to the  {\em Chandra} position
along \pks \/ line of sight by \citep{nic02}.
A similar shift ($+44$ m\AA ) is present in the 23.3 \AA \/ molecular 
oxygen 
instrumental feature  (see \S ~\ref{sec_instrulines}) 
when compared to the {\em Chandra} position by \citep{nic02}.
The systematic rigid redshift of the wavelength scale in the RGS-1 
spectrum of \pks \/ is larger than both the calibration uncertainty
of 8~m\AA\/ \citep[$1$~$\sigma$][]{den01} and the intrinsic $\sim 16$ m\AA \/
$1$~$\sigma$ uncertainty in the line centroid measurement.
For consistency with the {\em Chandra} measurements, we then decided 
to apply a constant shift of $-35$ m\AA \/ to all the RGS-1 
line positions reported in this paper. 
The RGS-2 correction factor is more difficult to estimate because  
the 23.5 \AA \/ feature
falls on the failed RGS-2 CCD~4.

\subsubsection{Instrumental lines}\label{sec_instrulines}

\begin{figure}
\plotone{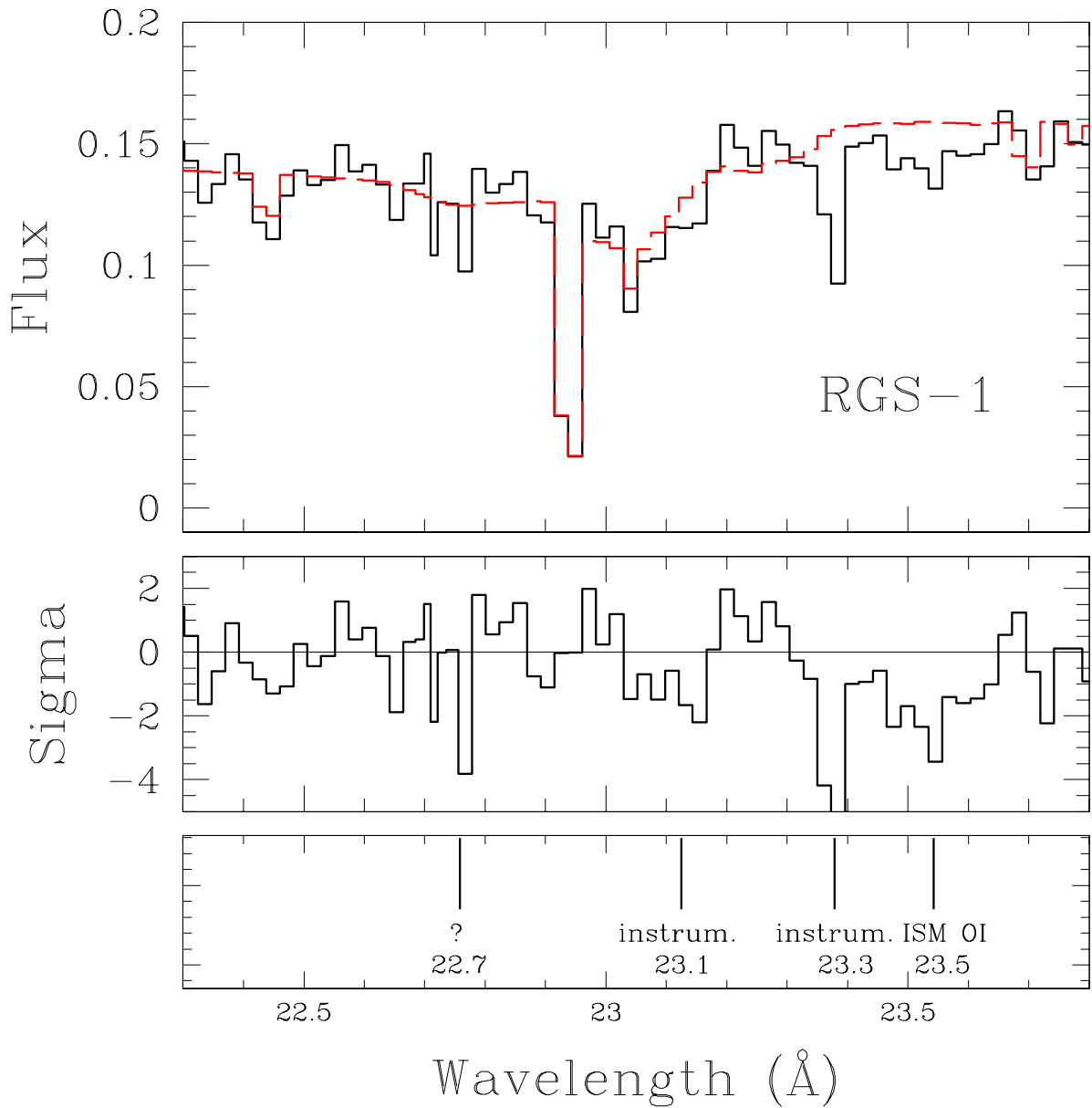}
\caption{Top panel: RGS-1 first order spectrum and best fit absorbed power law 
 model folded with the instrument response (dashed line).
The middle panel shows the residuals to the fit and the bottom panel shows the position of the absorption features. \label{fig3}}
\end{figure}

The two lines at  23.07 \AA \/ and at  23.35 \AA \/ (Fig.~\ref{fig3})
are likely to be  broadened and shifted  instrumental
molecular oxygen lines  \citep{dev03}. 
The 23.35 \AA \/ line is also present in {\em Chandra} spectra, while the 
23.07~\AA \/ line is a {\em XMM-Newton} feature.
These lines are also present in the {\em XMM-Newton} spectrum Sco X-1 
{\em XMM-Newton} \citep{dev03} and they are interpreted as instrumental 
shifted and broadened 1s$\rightarrow$2p lines of bound oxygen.
Since {\em XMM-Newton} detectors do not contain atomic oxygen 
in their structure, these lines are probably caused by oxides 
\citep[SiO$_2$, Al$_2$O$_3$ and the hygroscopic MgFe$_2$][see de Vries et al. 2003 for a detailed discussion on these features]{den02}.

We note that an absorption feature appears in the RGS-1 and in the RGS-2
 spectra  at 19.5 \AA \/ 
(Fig.~\ref{fig4}) with significance of 2.3 and 2.5 $\sigma$  respectively
and an EW of $\sim 8$ m\AA \/ ($\sim 0.3$~eV).
The same line is not present in the {\em Chandra} spectra of \pks \/ 
 \citep{nic02, fan02} but it is visible in the {\em XMM-Newton}  
spectrum of MRK~421 \citep{cag03} 
and therefore it is likely to have an instrumental origin.

\subsubsection{Lines of astrophysical interest}\label{sec_astrolines}

The lines of astrophysical interest are reported in Table~\ref{tab4}.
One of the strongest lines in our spectrum is 
the O{\sc vii} K$\alpha$ line at 21.6 \AA \/  (Figure~\ref{fig5} and Table~\ref{tab4}).
The O{\sc vii} K$\alpha$  profile shows two peaks at 21.5 and
21.6  \AA .
The wavelength separation between the peaks is $\sim 0.085$ \AA , 
larger than the resolution limit at this wavelength 
\citep[$\sim 0.066$ \AA ][]{dev03}, and corresponds to
$\sim 1180$ km s$^{-1}$. 
However we note that the significance of the splitting is statistically low,
in the sense that no improvement  of the fit with two Gaussian is found.
Moreover note  that the quoted velocity is 
more than a factor of three higher than the highest
 velocities observed by FUSE for the O{\sc vi} clouds \citep[e.g.][]{nic02b, sem02}.

\begin{figure}
\plottwo{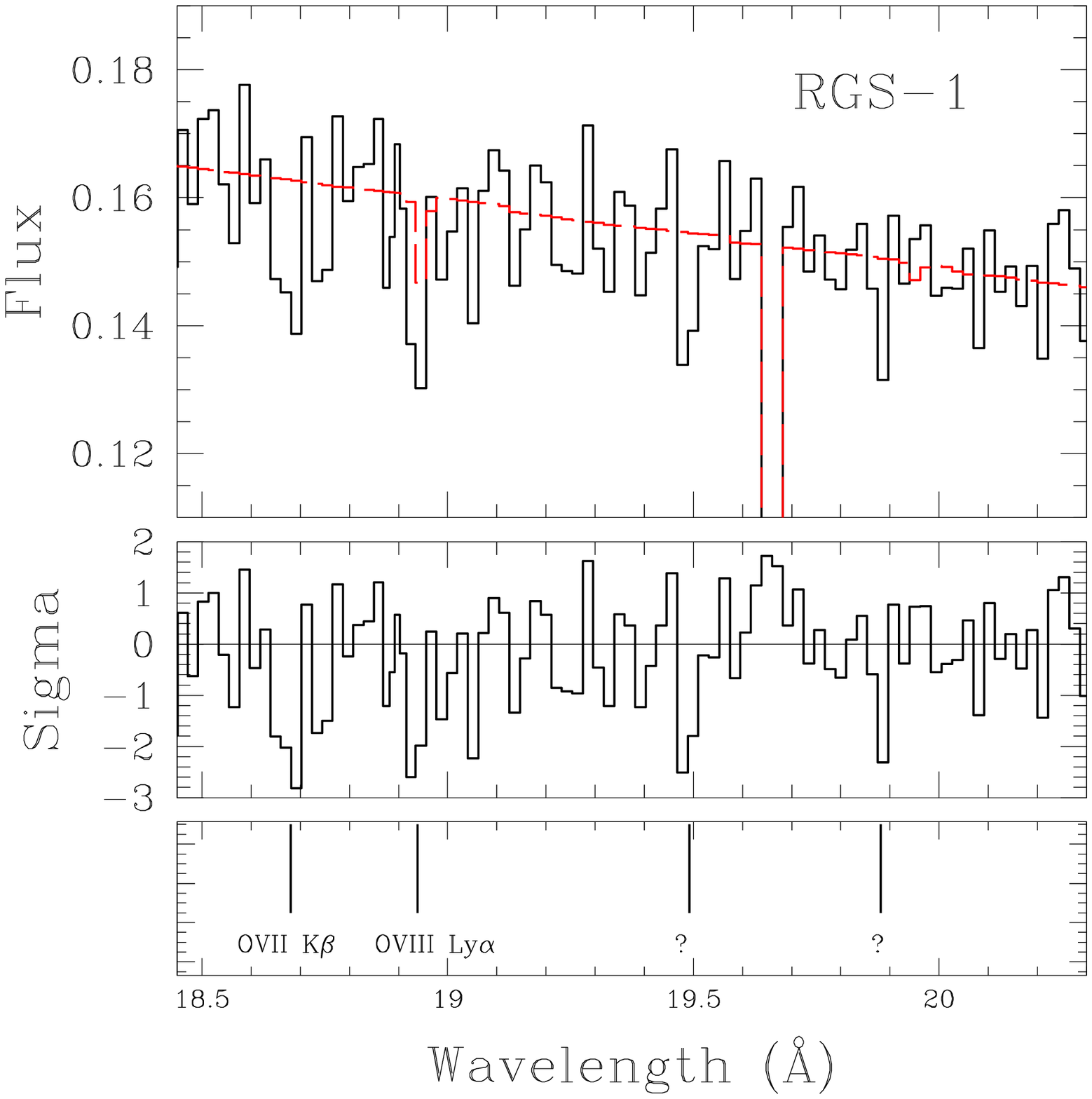}{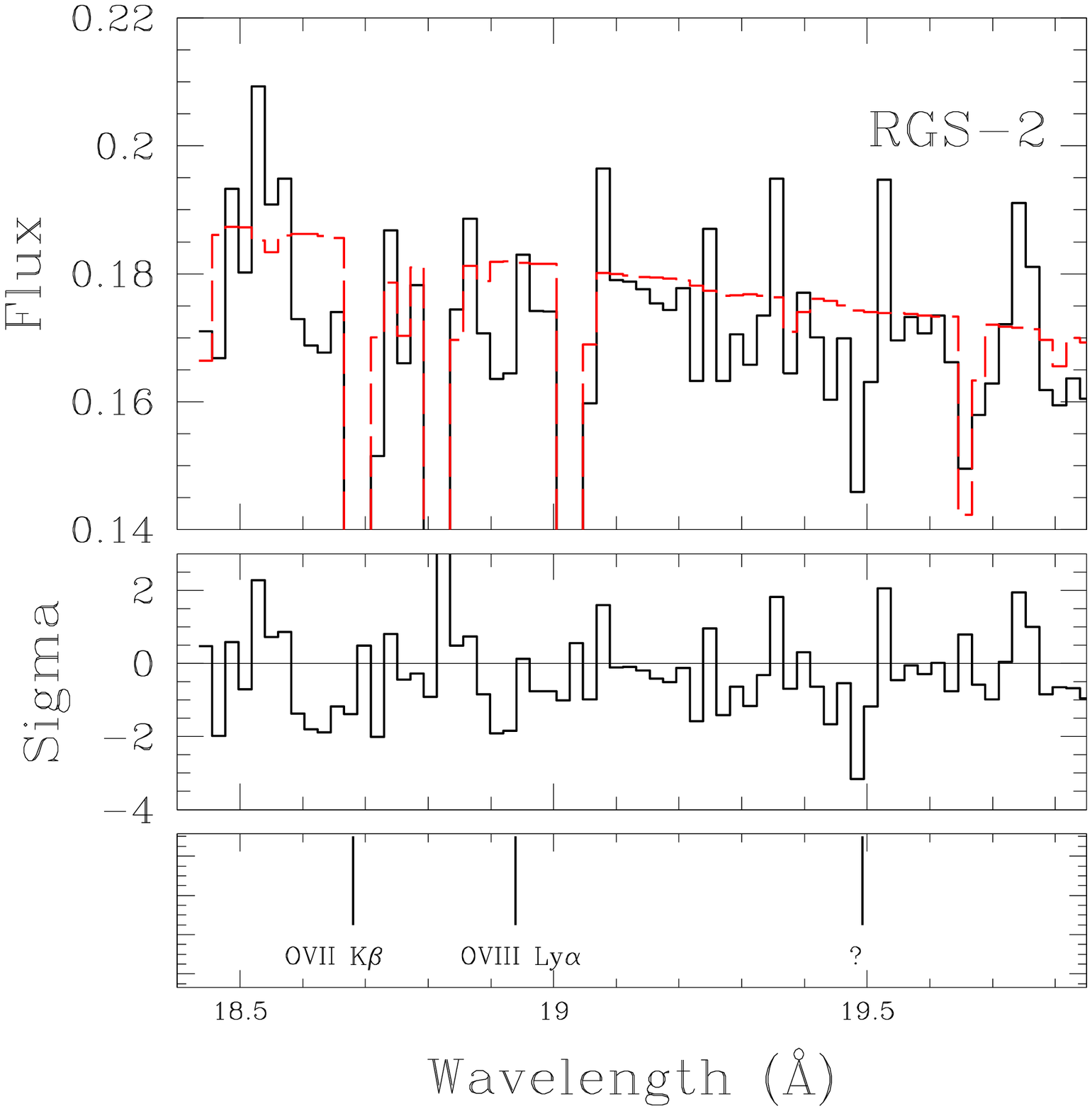}
\caption{Same as \ref{fig3} for RGS-1 (left) and RGS-2 (right)  first order  spectra in the 18.4 - 20.5 \AA \/ range. \label{fig4}}
\end{figure}

\begin{figure}
\plotone{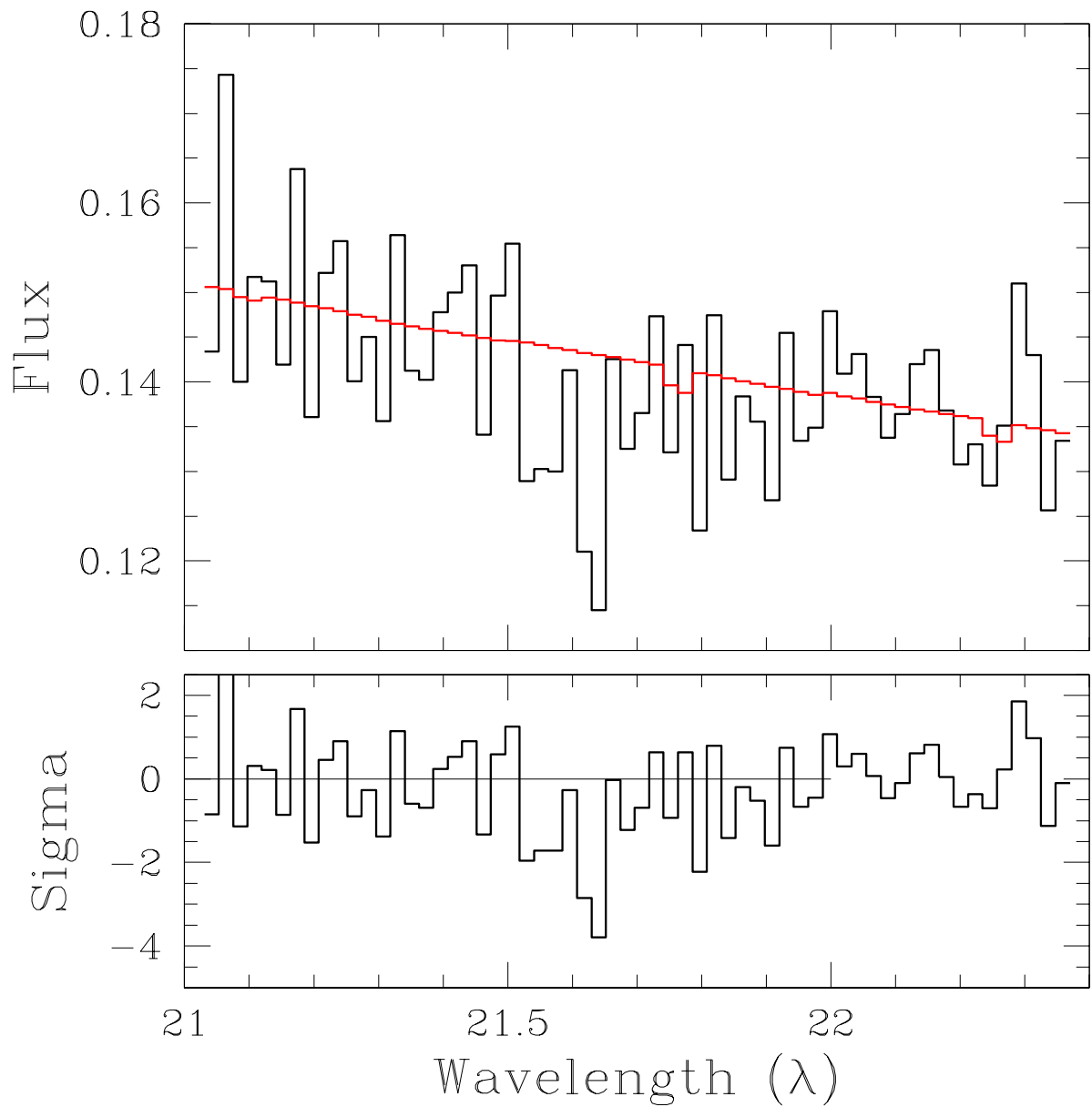}
\caption{Top panel: RGS-1 first order spectrum and best fit absorbed
 power law model folded through the instrument response in the 
21--22 \AA \/ region. Residuals to the fit are reported in the bottom panel.
\label{fig5}}
\end{figure}

The corresponding  O{\sc vii} K$\beta$,
is visible in RGS-1 at $\sim 18.64$ \AA \/ (Fig.~\ref{fig4} left)
but it overlaps with an effective area  feature in RGS-2
 (Fig.~\ref{fig4} right). 
Since a double-peaked profile is not detected for this line, the 
double peak appearance of the O{\sc vii} K$\alpha$ profile could be due 
to a  positive fluctuation within the line, consistently with the arguments given above.

The EW of the O{\sc vii} K$\alpha$ 
($19.50^{+7.89}_{-8.17}$ m\AA \/ or $0.517 \pm 0.210$ eV)
is consistent with the values presented in 
\citet{nic02} ($0.31^{+0.20}_{-0.16}$ eV) using {\em Chandra}
and in \citet{ras03} ($16.3 \pm 3.3$ m\AA ) using a coaddition of 
{\em XMM-Newton} calibration observations for a total of 265~ks.
As derived from the curve of growth presented in Fig.~4  of \citet{nic02},
the  O{\sc vii} K$\alpha$ EW is compatible with
an unsaturated line.
To check if any effective area feature could be present in the positions of 
the  O{\sc vii} lines, we extracted the RGS-1 spectra of the Crab nebula
and of another BL Lac object: H~1426$+$428.
No  O{\sc vii} K$\alpha$ is detected in the 46~ks long observation of the BL Lac; 
but a $\sim 20$\% feature at  18.65 \AA \/ is visible.
Similarly the Crab  spectrum shows a $\sim 10$\% drop in the same region.
Therefore the O{\sc vii} K$\beta$, detected in RGS-1 at $\sim 18.64$ \AA \/ 
(see Figure~\ref{fig4}, left panel), 
is likely to be affected by the presence of this effective area drop.
Since we cannot correct for this effective area feature, the 
line EW reported in Tab.~\ref{tab4} is to be considered as an upper limit only.
The EW ratio of O{\sc vii} K$\alpha$ 
and K$\beta$ limit is $>1.29^{+1.46}_{-0.76}$, consistent with 
\citet{nic02} value of $\sim 1.41$, and with the oscillator 
strength ratio of 4.77.

We do not detect any O{\sc viii} Ly$\alpha$ at 18.97 \AA . 
Fixing the line width to the value obtained for O{\sc vii} K$\alpha$, 
we find a 3 $\sigma$ upper limit of 12.61 m\AA \/ for the line EW (Tab.~\ref{tab4}).
We note however that the 3~$\sigma$ upper limit corresponds to a 
 O{\sc viii} Ly$\alpha$ normalization 11\% of the power-law value, 
comparable to the depth of a nearby (18.92 \AA ) effective area drop. 

At the redshift of the intervening concentration of galaxies ($z\sim 0.005$) 
we do not confirm the strong O{\sc viii} K$\alpha$ present in \pks \/  
{\em Chandra} spectrum of \citet{fan02} but not present 
in \citet{nic02} nor in \citet{ras03}.
The 20.02~\AA \/ line  EW in \citet{fan02} is $\sim 14^{+7.3}_{-5.6}$
m\AA , comparable to the 21.6 \AA \/ O{\sc vii} K$\alpha$ EW and to the 
 18.6 \AA \/ O{\sc vii} K$\beta$ in this spectrum.
Since we detect the O{\sc vii} K$\alpha$  \citep[with EW consistent with][]{fan02},
 the lack of detection of the 20.02 \AA \/
feature is unlikely to be the result of insufficient statistics in our 
continuum. An upper limit can be set on the equivalent width of a putative 
absorption line at 20.02 \AA \/ of 14.0 m\AA , at 3 $\sigma$ confidence 
level, i.e. the value observed by \cite{fan02}.

It is not possible to determine if any Ne{\sc ix} absorption 
is present because RGS-2 only is active in this wavelength range and
 it presents a $\sim 21$\% drop in the effective area at $\sim 13.45$~\AA .
We  detect at 1.4 $\sigma$ at $\sim 22.72 \pm 0.03$ \AA \/ with EW 
of $\sim 6$ m\AA \/ the 22.77 \AA \/ line 
already seen  in \pks \/ and also in MRK~421 RGS spectra by \citet{dev03}.
The best fit parameters to the  23.5 \AA \/ interstellar O{\sc i}  
1s$\rightarrow$2p line, discussed in \S ~\ref{sec_shift}, 
are reported in Table~\ref{tab4}.

\section{Discussion}
\label{sec_disc}

At $z \sim 0$ we confirm the existence of a O{\sc vii} K$\alpha$ 
absorption feature, but we do not detect
features of higher ionization elements  (O{\sc viii}, Ne{\sc ix} and 
Ne{\sc x}) seen by {\em Chandra} \citep{nic02}.
This lack of detection is likely to be related to the limited number 
of photons (e.g. $\sim 15300$ photons between 19.5 and 20.5 \AA \/, 
i.e. $\sim 100$ photons per 0.0065 \AA \/ resolution element)
collected in the {\em XMM-Newton} observation, when the source 
was $\sim 2-3$ times fainter than in \citet{nic02}.
The  O{\sc vii} K$\alpha$ appears to be unsaturated and the curve of 
growth presented in  \citep{nic02} suggests an
 O{\sc vii} column density of $\sim 10^{16}$ cm$^{-2}$.
If the weak evidence of a double peak in the profile of the O{\sc vii} 
K$\alpha$ line  is considered real,
such shape may indicate that we are starting to detect the different 
sheets of the local filament.

The $3$~$\sigma$ upper limit on the O{\sc viii} Ly$\alpha$ line EW, 
computed fixing the line FWHM to the  O{\sc vii} K$\alpha$ FWHM value, 
 is $12.61$ m\AA , consistent with the value of 
$9.0^{+2.6}_{-2.7}$ m\AA \/ found by \citet{ras03} in the RGS spectrum of \pks .
Under the assumption of unsaturated lines the EW ratios between different 
ions of the same element depend on the gas temperature and density.
Using  the ratio between the upper limit on the EW of  
O~{\sc viii} Ly$\alpha$
to the  O{\sc vii} K$\alpha$ (best fit value $-2$~$\sigma$), 
we obtain an upper limit on the 
gas temperature of $\sim 2.5-3.5 \times 10^6$~K, for a  gas density 
of $10^{-6}$- 1 atom cm$^{-3}$ \citep[see Fig.~5 in][]{nic02}.
These values are consistent with the temperature range predicted for the 
WHIM.

It is not possible to discriminate with the present data whether 
the O{\sc vii} features are 
produced by a WHIM, as proposed by \citet{nic02} or by radiatively cooling 
gas within our Galaxy, as proposes by \citet{hec02}.
In the {\em FUSE} spectrum  of \pks \/ \citet{nic02} found two unsaturated O{\sc vi} absorption lines: a narrow component (FWHM = $106 \pm 9$ km s$^{-1}$) at $cz = 36 \pm 6$ km s$^{-1}$ with EW = $(2.1 \pm 0.2) \times 10^{-3}$~eV, probably related to a cloud in the  Galactic disk,
and a broad component (FWHM = $158 \pm 26$ km s$^{-1}$) at $cz = -135 \pm 14$ km s$^{-1}$ with EW = $(1.6 \pm 0.4) \times 10^{-3}$~eV possibly associated with a WHIM filament.
Using our O{\sc vii} K$\alpha$ EW and the EW of the O{\sc vi} broad component from \citet{nic02} we derive a temperature range of $\sim (4.5 -200) \times 10^{5}$~K for a gas density of $\sim 10^{-6}$ cm$^{-3}$, which restricts to $\sim (4.5 - 25) \times 10^{5}$~K once the upper limit from the O{\sc viii} is considered.
We note, however,  that the  O{\sc vii} absorption features at local redshift 
have been found in the spectra of other 4 AGNs (3C~273 \citet{fan03}; 
 MRK~421 \citet{nic01,cag03}; NGC~4593 \citet{mck03} 
and NGC~3783 \citet{kas02}) and might 
be associated to  the high velocity O{\sc vi} lines 
(sampling a gas at $T\sim 10^5-10^6$ K) 
recently seen in the spectra of bright AGNs by FUSE \citep{sem02, nic02b}.
If this is the case, the ``local''  O{\sc vii} K$\alpha$ would indicate 
the presence of WHIM within the Local Group  \citep{nic02b}.

The line at 23.5 \AA \/ is  the interstellar O{\sc i} 
1s$\rightarrow$2p transition  \citep[e.g.][and references therein]{mcl98}.
Assuming a neutral ISM gas, no consistent velocity
broadening and unsaturated lines, the comparison of the measured  EW 
with the curve of growth derived using \citet{sto97} cross sections implies 
$\sim 2 \times 10^{16}$ O atoms cm$^{-2}$.
The H absorbing column density in \pks \/ direction is 
$1.36 \times 10^{20}$ cm$^{-2}$ \citep{loc95}, which brings to  
a O/H ratio of $1.4 \times 10^{-4}$.
This value is  $\sim 6$ times smaller than in the Solar system 
\citep[$8.5 \times 10^{-4}$][]{and89}, but consistent with the ISM elemental 
abundances in the Magellanic Clouds \citep[e.g.][]{rus90}.
Assuming that the line EW is enhanced by velocity broadening would 
further decrease the O/H ratio.
If we adopt the theoretical cross-sections of \citep{mcl98},
as in \citet[][(see their Fig.~3)]{dev03}, 
we derive $\sim 7 \times 10^{16}$ O atoms cm$^{-2}$,
i.e. a O/H ratio of $\sim 5 \times 10^{-4}$, still
$\sim 1.5$ times lower than the Solar value.
We note that \citet{dev03}  O{\sc i} 
1s$\rightarrow$2p EW in \pks \/ direction, 
derived from a coaddition of RGS observations for a total of 346~ks, 
is $15 \pm 3$ m\AA \/ consistent with
our measurement ($17.57^{+1.21}_{-5.81}$ m\AA ) and the line curve of 
growth presented in the paper takes saturation effects into account, 
but still there is indication of a subsolar O/H ratio.

We do not confirm the O{\sc viii} K$\alpha$ line by \citet{fan02}.
Note that the lack of detection of WHIM outside our Local Group is consistent
with the expectations of theoretical models. 
In fact only one O{\sc viii} absorption line with EW$>3$ m\AA \/ 
(i.e. detectable by the present X-ray satellites) in the spectrum of a  random background 
source at $z\sim 0.3$ is predicted \citep[e.g.][]{hel98}. 
The non-detection of WHIM up to $z =0.116$  therefore does not pose any problem.

Our {\em XMM-Newton} \pks \/ spectrum shows a hint 
($1.4$~$\sigma$ detection; EW $\sim 5.56$~m\AA ) of a 22.7 \AA \/ feature, 
seen by 
\citet{dev03} in the  {\em XMM-Newton} spectra of \pks \/ and MRK~421, 
and in the {\em Chandra} spectrum of \pks .
Since they do not find evidence of such line in the {\em XMM-Newton} 
spectra of  Sco~X-1 and 4U~0614$+$091,  
they  exclude a possible instrumental origin of the line and tentatively 
identify it with O{\sc iv} absorption from a local WHIM filament.
Higher statistics and other time spaced observations are needed to 
investigate the reality of the 22.7 \AA \/ absorption feature.

\acknowledgments

This work made use of the  NASA/IPAC
Extragalactic Database (NED) which is operated 
by the Jet Propulsion Laboratory,
CalTech, under contract with the NASA. 
I.C. thanks Matteo Guainazzi, Maria Santos-Lleo and the XMM helpdesk for the
precious help during the reduction process and 
Francesco Haardt for useful scientific discussion.
I.C. acknowledges a C.N.A.A. fellowship.
I.C. is infinitely grateful to Vinay Kashyap for the discussion, 
suggestions and help in the effective area and response matrix related issues, 
that brougth to a succesful coaddition of the spectra using the PINTofALE 
interactive data language software suite \citep{kas00}.

\clearpage

\clearpage

\clearpage 

\clearpage 

\begin{deluxetable}{c cccc}
\tablecaption{{\em XMM-Newton} RGS observations of \pks \/ in revolution 174. \label{tab1}}
\tablewidth{0pt}
\tablehead{
\colhead{Obs. Id.}
& \colhead{Start\tablenotemark{a}}   &\colhead{Stop\tablenotemark{a}} &\colhead{Total Exposure\tablenotemark{b}} &\colhead{Net Exposure\tablenotemark{c}}
}
\startdata
0080940401 &19, 15:36:56 &19, 18:30:22	&10 &9  \\
0080940101 &19, 18:38:20 &20, 11:26:51	&60 &58 \\
0080940301 &20, 12:53:01 &21, 05:56:27	&61 &44 \\
0080940501 &21, 05:58:37 &21, 07:28:45  &5  &0  \\
\enddata
\tablenotetext{a}{days of November 2000, time is Terrestrial Time. RGS~1 and RGS~2 were operated simultaneously and have start and stop time differing of few seconds only.}
\tablenotetext{b}{Total RGS on time in ks.}
\tablenotetext{c}{In ks after the high particle background times rejection.}
\end{deluxetable}

\clearpage 

\begin{deluxetable}{c|ccc|ccc}
\tablecaption{Fit to the RGS-1 and RGS-2 spectra with an ansorbed power law model with absorption fixed to the Galactic value ($N_H =1.36 \times 10^{20}$ cm$^{-2}$).\label{tab2}}
\tablewidth{0pt}
\tablehead{
\colhead{Obs. Id.} &\multicolumn{3}{c}{RGS~1} &\multicolumn{3}{c}{RGS~2}\\
& \colhead{$\Gamma$}   &\colhead{Norm.}\tablenotemark{a} &\colhead{Flux}\tablenotemark{b} 
& \colhead{$\Gamma$}   &\colhead{Norm.}\tablenotemark{a} &\colhead{Flux}\tablenotemark{b}\\
&	& &(0.5-2.0 keV)
&	& &(0.5-2.0 keV)\\
}
\startdata
0080940401 	&$2.438 \pm 0.013$	&$4.08 \pm 0.03$ 	&$8.76 \pm 0.07$	
		&$2.426 \pm 0.012$	&$3.98 \pm 0.03$	&$8.53 \pm 0.06$\\
0080940101 	&$2.488 \pm 0.006$	&$3.25 \pm 0.01$	&$7.01 \pm 0.03$	
		&$2.469 \pm 0.005$ 	&$3.17 \pm 0.01$	&$6.82 \pm 0.02$\\
0080940301 	&$2.573 \pm 0.008$	&$2.58 \pm 0.01$  	&$5.59 \pm 0.03$	
		&$2.555 \pm 0.007$	&$2.49 \pm 0.01$	&$5.38 \pm 0.02$\\
\hline\\
Total           &$2.491 \pm 0.005$      &$3.06 \pm 0.01$        &$6.59 \pm 0.02$                &$2.472 \pm 0.005$      &$2.97 \pm 0.01$        &$6.40 \pm 0.02$\\
\enddata
\tablenotetext{a}{at 1 keV in units of $10^{-2}$ ph cm$^{-2}$ s$^{-1}$ keV$^{-1}$}
\tablenotetext{b}{In units of $10^{-11}$ erg cm$^{-2}$ s$^{-1}$}
\end{deluxetable}

\clearpage

\begin{landscape}
\begin{table}
\begin{center}
\caption{Best-fitting RGS-1 absorption line parameters and $1 \sigma$ errors derived from a local simultaneous fit (see text for details). \label{tab4}}
\begin{tabular}{cccccc}
\tableline\tableline
Line ID
&$\lambda$ &$cz$\tablenotemark{a} &FWHM &EW\tablenotemark{b} &$\sigma$\\
&(\AA ) &(km s$^{-1}$) &($10^{-2}$ \AA ) &(m\AA ) &\\
\tableline
O{\sc vii} K$\beta$\tablenotemark{c,d}
&$18.646 \pm 0.10$	     &$282^{+173}_{-161}$  &$0.003^{+3.30}_{-0.003}$ &$15.13^{+6.12}_{-5.15}$ &3.7\\
O{\sc viii} Ly$\alpha$\tablenotemark{e}
&$18.967$ 	&$0$ &$0.12$ &$<12.61$ &0.7\\
O{\sc vii} K$\alpha$	
&$21.586^{+0.017}_{-0.016}$ 	&$-223^{+222}_{-236}$ &$5.97^{+3.94}_{-5.97}$ &$19.50^{+7.89}_{-8.17}$ &4.5\\
molecular O{\sc i}	
&$23.066^{+0.038}_{-0.031}$     &--  &$4.16^{+3.04}_{-4.16}$  &$10.20^{+6.80}_{-5.81}$ &2.2\\
molecular O{\sc i}\tablenotemark{c}	
&$23.350^{+0.002}_{-0.007}$ 
  &$115^{+32}_{-92}$ &$0.003^{+1.298}_{-0.003}$ &$22.58^{+6.62}_{-5.54}$ &5.5\\
O{\sc i} K$\alpha$ 
&$23.510 \pm 0.015$     &$16^{+192}_{-189}$ &$5.20^{+2.78}_{-3.38}$ &$17.57^{+1.21}_{-5.81}$ &4.6\\
\tableline
O{\sc viii} K$\alpha$\tablenotemark{f}	
&20.02	      &16624 &6.49 &$1.54^{+4.17}_{-1.54}$ &0.4\\
\tableline
\end{tabular}
\tablenotetext{a}{Note that
 we applied a $-35$ m\AA \/ shift to RGS-1 wavelength calibration in order to match \citet{nic02} 
position for the interstellar O{\sc i}  1s$\rightarrow$2p  line.} 
\tablenotetext{b}{The EW is computed as the mean of the EW of the line in the three 
spectra as a result of a simultaneous fit. The errors are computed as the mean of the three 
values obtained allowing the line and the power laws normalization to vary within their  1 $\sigma$ limits}
\tablenotetext{c}{Results obtained excluding the first observation  (ID 0080940401) because of the presence of an effective area feature on the line position.}
\tablenotetext{d}{The line parameters are likely to be affected by calibration uncertainties. The line EW and FWHM can be regarded as an upper limits only.}
\tablenotetext{e}{The line position and line width are fixed to the expected value and to the O{\sc vii} K$\alpha$ value respectively. The EW corresponds to the 3~$\sigma$ upper limit.}
\tablenotetext{f}{The line position and the line width are fixed to the value and to the 90\% upper limit in \citet{fan02} respectively.}
\end{center}
\end{table}
\end{landscape}

\end{document}